# Task Scheduling of Multiple Agile Satellites with Transition Time and Stereo Imaging Constraints

**Junhong Kim[1], Jaemyung Ahn[2], Han-Lim Choi[3]**

*Korea Advanced Institute of Science and Technology (KAIST), Daejeon 34141, Republic of Korea*

**Doo-Hyun Cho[4]**

*Samsung Electronics, 1-1 Samsungjeonja-ro, Hwaseong, Gyeonggi 18448, Republic of Korea*

## Abstract

This paper proposes a framework for scheduling the observation and download tasks of multiple agile satellites with practical considerations such as attitude transition time, onboard data capacity, and stereoscopic image acquisition. A mixed integer linear programming (MILP) formulation for optimal scheduling that can address these practical considerations is introduced. A heuristic algorithm to obtain a near-optimal solution of the formulated MILP based on the time windows pruning procedure is proposed. A comprehensive case study demonstrating the validity of the proposed formulation and heuristic is presented.

**Keywords**: multiple satellite operation; task scheduling; stereoscopic imaging; mixed integer linear programming (MILP); heuristics

## Nomenclature

### Indices / Index Sets

$v$ / **V**     Observation task

$d$ / **D**     Download task

$s$ / **S**     Satellite

$g$ / **G**     Ground station

---

[1]  Graduate Research Assistant, Department of Aerospace Engineering, 291 Daehak-ro, Yuseong, Daejeon, Republic of Korea

[2]  Associate Professor, Department of Aerospace Engineering, 291 Daehak-ro, Yuseong, Daejeon, Republic of Korea; email: jaemyung.ahn@kaist.ac.kr (Corresponding Author), AIAA Member

[3]  Associate Professor, Department of Aerospace Engineering, 291 Daehak-ro, Yuseong, Daejeon, Republic of Korea

[4]  Staff Engineer, Mechatronics R&D Center, Samsung Electronics, 1-1 Samsungjeonja-ro, Hwaseong, Gyeonngi, Republic of Korea

k / $\mathbf{K}_{vs}^{\mathbf{V}}$     Observation time window (OTW) associated with task $v$ and satellite $s$

l / $\mathbf{L}_{ds}^{\mathbf{D}}$     Download time window (DTW) associated with task $d$ and satellite $s$

## Decision Variables

$x_v$ ( $v \in \mathbf{V}$ )     Selection of observation task $v$

$x_{v_s}$ ( $v \in \mathbf{V}$ )     Selection of stereoscopic observation task $v$

$x_{vs}$ ( $v \in \mathbf{V}, s \in \mathbf{S}$ )     Assignment of observation task v to satellite $s$

$x_{vsk}$ ( $v \in \mathbf{V}, s \in \mathbf{S}, k \in \mathbf{K}_{vs}^{\mathbf{V}}$ )     Selection of OTW k for task $v$ conducted by $s$

$t_{vsk}$ ( $v \in \mathbf{V}, s \in \mathbf{S}, k \in \mathbf{K}_{vs}^{\mathbf{V}}$ )     Start time of observation task $v$ conducted by satellite $s$ at OTW $k$

$t_{dsl}^{a} / t_{dsl}^{b}$ ( $d \in \mathbf{D}, s \in \mathbf{S}, l \in \mathbf{L}_{ds}^{\mathbf{D}}$ )     Start/end time of download task $d$ conducted by satellite $s$ at DTW $l$

$\theta_{vsk}$ ( $v \in \mathbf{V}, s \in \mathbf{S}, k \in \mathbf{K}_{vs}^{\mathbf{V}}$ )     Pitch angle associated with observation $v$ for satellite $s$ at OTW $k$

## Parameters

$T_H$  [s]     Scheduling Time Horizon

$w_v$     Priority of observation task $v$

$\phi_{vsk}$  [rad]     Roll angle required for $k^{th}$ OTW associated with task $v$ and satellite $s$ ( $v \in \mathbf{V}, s \in \mathbf{S}, k \in \mathbf{K}_{vs}^{V}$ )

$\alpha_v$  [rad]     Maximum roll/pitch angles requested by the user ( $v \in \mathbf{V}$ )

$\phi_s^M / \theta_s^M$  [rad]     Maximum roll/pitch angles obtainable by satellite s ( $s \in \mathbf{S}$ )

$I_s$  [Byte]     Initial data of satellite $s$ before scheduling ( $s \in \mathbf{S}$ )

$r_s$  [rad/s]     Slew rate of satellite $s$ ( $s \in \mathbf{S}$ )

$T_{vs}^{p}$  [s]     Observation process time associated with task $v$ and satellite $s$ ( $v \in \mathbf{V}, s \in \mathbf{S}$ )

$T_s^{r_s}$  [s]     Stabilization time of a satellite $s$ after attitude maneuver ( $s \in \mathbf{S}$ )

$T_g^{r_g} / T_s^{r_s}$  [s]     Preparation time for ground station $g$ ( $g \in \mathbf{G}$ ) / satellite $s$ ( $s \in \mathbf{S}$ )

$\beta_v$  [rad]     Required angle difference by the user for stereoscopic tasks ( $v \in \mathbf{V}$ )

$u_s$  [Byte]     Max data capacity for satellite s ( $s \in \mathbf{S}$ )

$\gamma_s$  [Byte/s]     Data download rate for satellite $s$ ( $s \in \mathbf{S}$ )

$\zeta_s$  [Byte/s]     Data acquisition rate for satellite $s$ ( $s \in \mathbf{S}$ )

$\lambda$     Time window clustering parameter

# I. Introduction

Satellite imaging with optical cameras or radars has many advantages over other methods such as aerial imaging and has been used for various purposes such as environmental, meteorological, and three-dimensional mapping missions [1-2]. Usually the number of requests that a satellite receives per day generally exceeds its operational capability. A typical Earth observation satellite has up to 10 opportunities to acquire the image of a target every day, while the number of requested missions can be hundreds [3-5]. This situation necessitates the selection and scheduling of imaging tasks to maximize the benefits obtainable by operating satellites [6-12].

The introduction of agile earth observation satellites (AEOSs) provides opportunities for highly efficient image acquirement potential [13-14]. An agile satellite can utilize attitude maneuvers actively for imaging targets whose coverage was intractable in the past. However, the agility entails the coupling between the observation start time and the pitch angle of the satellite as well. Studies on the scheduling of AEOS have primarily addressed its mathematical formulation [15-16] and the algorithms to obtain the schedule solutions, most of which are suboptimal [2, 17-26].

High attention has been paid on the imaging task scheduling with multiple satellites relatively recently [3, 25, 27-35]. The review of past studies on scheduling of agile satellite imaging led the authors to identify the opportunity primarily for improving the quality of a schedule solution. For example, an effective schedule should consider observation and download simultaneously to reflect their coupling. In addition, satellite scheduling is especially sensitive to the modeling methodology for the transition time between tasks. Some studies used over-simplified operational assumptions such as constant attitude change time, which, in practice, varies with the task start time. Some other studies adopted piecewise linear functions for better solution quality.

This study modifies the authors' previous work [35] and introduces an improved framework for task scheduling of a group of agile satellites taking the images of targets located on the Earth surface with additional consideration of stereo imaging and time-dependent attitude transition time. The mathematical formulation for the scheduling considers the priority of the imaging tasks in its objective function and reflects practically important issues (e.g., the coupling between the task time and required attitude maneuver, data capacity, and stereoscopic imaging) as explicit constraints.

Three major contributions of this study are as follows. First, we propose a mixed-integer linear programming (MILP) formulation for practical scheduling of imaging tasks for multiple satellites considering their agility and associated complications. Some past studies looked into the MILP based methods for optimal satellite task scheduling (e.g., [36-39]). However, very few published literature presented the formulation explicitly considering practical issues such as the coupling between the task start time and required satellite attitude, and constraints on simultaneous observation and download scheduling, which are addressed in this study. Second, a new MILP-based heuristic for the scheduling problem is developed to significantly reduce its computation time. Due to the inherent complexity of agile satellite scheduling, obtaining exact solutions for large problem instances (e.g., with hundreds of tasks) requires very long scheduling time unsuitable for runtime applications. To address this issue, we introduce a priority-based time-window pruning technique that can reduce the search space of the original MILP formulation and provide near-optimal solutions within reasonable scheduling time. Finally, we present a comprehensive case study that demonstrates the validity of the proposed scheduling algorithm. The performances (accuracy and computation time) of the proposed MILP formulation and heuristic are systematically presented for eight realistic Earth imaging scenarios.

## II.    Problem Description

Consider a group of satellites operating to fulfill the imaging requests of users by conducting the *observation* and *download* tasks. The observation and download *time windows* are associated with the two task types, respectively. If the task is stereoscopic, the satellite should observe the target twice under a certain geometric condition for its completion. Each observation task is associated with a profit value, which depends on the importance of the task. The time it takes to process the observation task at a given target is determined (by users) before the scheduling starts.

The profit is obtainable when the satellite successfully completes download of the acquired image data to a ground station. The objective of the problem is to maximize the sum of profits obtained by the satellites. The satellite can maneuver in both roll (left and right sides of the ground track) and pitch (forward and backward of the ground track) directions, as Figure 1 illustrates.

### A.   Observation Task

Observation requests submitted by users are inputs of the problem. The following four parameters define an observation request: 1) target type, 2) observation time window, 3) task priority, and 4) image type [40]. The target type can be either a spot or a polygon as described in Figure 2. While a single observation strip can cover a spot target, a polygon target requires multiple observation strips. In this paper, we assume that a set of multiple spot targets can approximate a polygon target. An observation time window specifies the start and end times within which the satellite can acquire the target image. An observation conducted outside of the window cannot obtain the profit assigned to the task. The task priority represents the importance of the task and determines the task's profit value, whose summation is the objective function of the problem to maximize. The target image, which is the final attribute used to define the task, can be *monoscopic* or *stereoscopic*. In addition, an observation task may require *optical* or *radar* equipment. Sunlight or cloud coverage may affect the observation possibility of optical observations, whereas radar observations are unaffected.

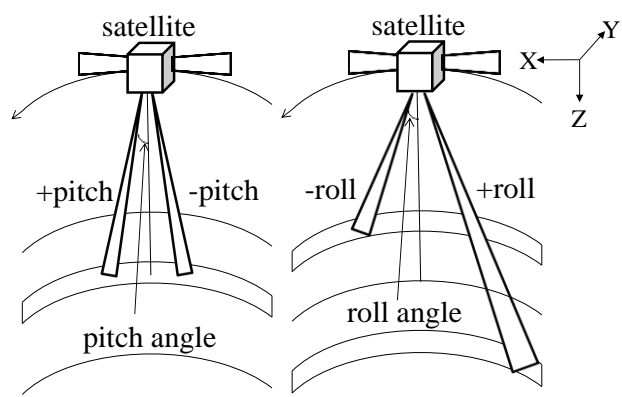

**Figure 1. Pitch (left) and roll (right) angles of a satellite**

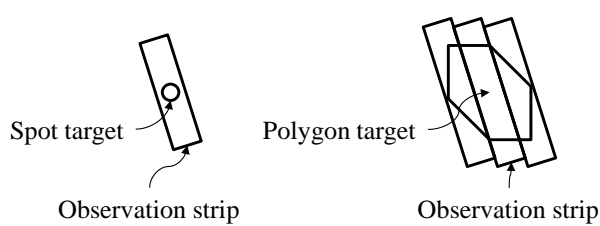

**Figure 2. Target type – spot target (left) and polygon target (right)**

### B. Observation Time Window

The observation time window is <span style="color:red">determined by considering</span> the limits of maneuvering angles (pitch angle, in particular). The start and end time points of the window correspond to the minimum and maximum pitch angles that a satellite is allowed to maneuver ( $-\theta_{max}$ / $+\theta_{max}$ ), respectively. Figure 3 shows the geometric relationship between the minimum and maximum pitch angles and the observation time window for a given target. Large attitude maneuver, which increases the length of the observation time window, may degrade the resolution of the acquired image.

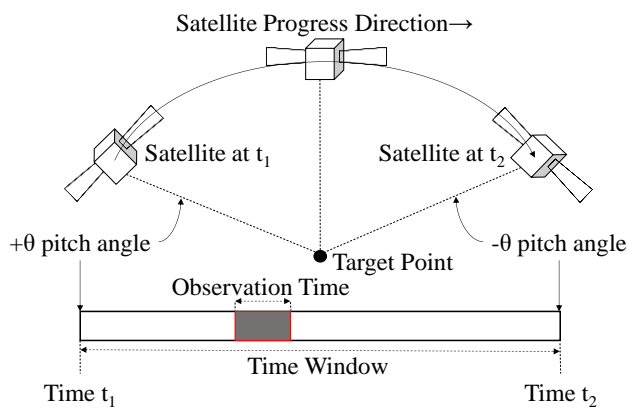

**Figure 3. Observation time windows (OTWs) and actual observation time**

### C. Download Time Window

A satellite can download the acquired image data when it has access to at least one of the ground stations. This accessibility condition imposes a constraint on download time windows. Unlike the observation task using an optical equipment, download task does not necessarily require direct sunlight for mission fulfillment. Therefore, satellites can download data regardless of sunlight status while observation tasks using optical equipment are feasible only during sunlight zones.

### D. Transition Time

Two transition time types – the first type is the time between observation tasks of a satellite and the second is the time between download tasks of a ground station – are considered. The second type, which is relatively simple to analyze, is the time required by a ground station between two consecutive download tasks.

After completing an observation task, a satellite should conduct the pitch and/or roll maneuver in preparation for the next observation, and the time required for this attitude change is the first type of transition time. Note that, as illustrated in Figure 3, the roll and pitch angles required for target pointing depend on the observation time – i.e., time dependent. Figure 4 presents sample profiles of required attitude angles during a typical time window (altitude: 700 *km*, field of view: 30 *deg*). This study models the pitch angle as a linear function of time and the roll angle as a constant during an observation time windows. Note that a recent study pointed out that this approach produces higher quality solutions than assuming a constant or piecewise linear transition time [41].

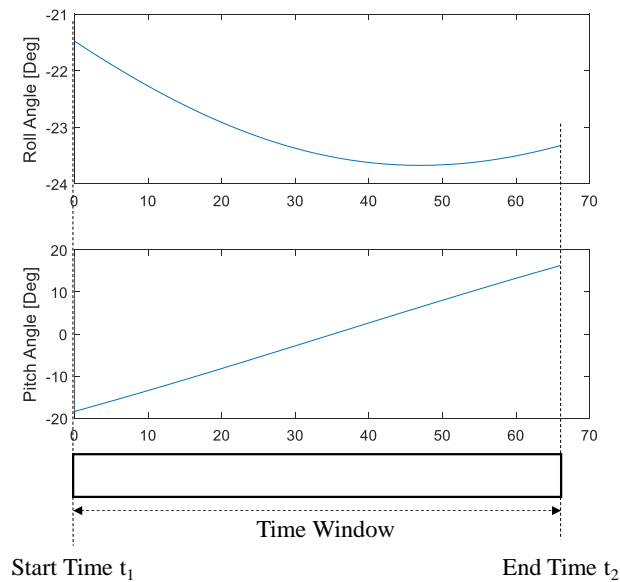

**Figure 4. Sample profiles of required roll and pitch angles for target pointing during a time window**

### E. Data Capacity

The onboard storage capacity of as satellite is limited and appropriate download scheduling between observation tasks is required. We also considered the initial data a satellite may possess. Near the end of a scheduling horizon, a satellite often cannot send its observation data to the ground station since there is no available time window but still can conduct observation tasks. In this study, we consider the remaining data as the initial data stored onboard at the next scheduling horizon.

### F. Satellite and Ground Station Task Overlap

We considered three different types of overlap between tasks. First, a satellite cannot perform two observation tasks simultaneously. Second, a ground station's radar typically cannot handle download tasks associated with two or more satellites at the same time due to constraints on the equipment and/or workforce. Third, synchronous download and observation of a satellite is allowed (the synchronous mode) [3].

### G. Time Window Requirement and Stereo Image Acquirement

An observation task or a download task must occur within a given time window. We assume that a monoscopic task requires a single observation and a stereoscopic task requires two observations. The two observations for a stereoscopic task (with different viewing angles) can take place 1) within an observation time window, or 2) in two separate time windows associated with a single satellite or two distinct satellites. Figure 5 presents an example of observation and download sequence in a scheduling horizon considering two satellites. The white boxes represent the observation time windows and download time windows. The red boxes are the task times for each observation task and the blue boxes are the task times for each download tasks, respectively. For cases where transition times between two different tasks are required, the time is shown in gray boxes.

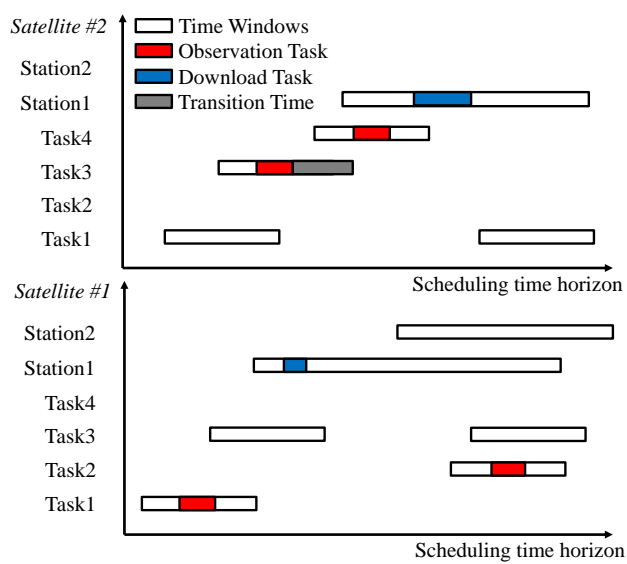

**Figure 5. Observation and download task sequence example**

## III.    Problem Formulation and MILP-based Heuristic

This section presents the MILP formulation for task scheduling of heterogeneous, multiple, and agile satellites with stereo imaging and other practical constraints and proposes a heuristic to efficiently obtain the near-optimal solution of the MILP.

### A.  MILP Formulation

A MILP formulation for an optimal task scheduling of multiple, heterogeneous, and agile satellites is presented as follows.

**[P: Optimal multi-heterogeneous agile satellites scheduling with stereo imaging]**

$$J = \max_{\mathbf{x}, \mathbf{t}, \mathbf{\theta}} \left( \sum_{v \in \mathbf{V}} (w_v \cdot x_v) \right) \tag{1}$$

subject to

$$x_{vs} = \sum_{k \in \mathbf{K}_{vs}^{\mathbf{V}}} x_{vsk} , \quad x_v = \sum_{s \in \mathbf{S}} x_{vs} \tag{2}$$

$$x_v \leq 1 \quad \text{(Active only for monoscopic tasks)} \tag{3}$$

$$x_v = x_{v_s} , \quad x_v \leq 1 \quad \text{(Active only for stereoscopic tasks)} \tag{4}$$

$$T_{vsk}^a \leq t_{vsk} \leq T_{vsk}^b \quad \text{if} \quad x_{vsk} = 1 \tag{5}$$

$$T_{dsl}^a \leq (t_{dsl}^a , t_{dsl}^b) \leq T_{dsl}^b \tag{6}$$

$$\begin{aligned} t_{vsk} + T_{vs}^p + T_s^{r_s} + \left| \phi_{vsk} - \phi_{v'sk'} \right| / r_s + \left| \theta_{vsk} - \theta_{v'sk'} \right| / r_s \leq t_{v'sk'} \\ \text{or} \qquad\qquad\qquad\qquad \text{if} \quad x_{vsk} , x_{v'sk'} = 1 \\ t_{vsk'} + T_{v's}^p + T_s^{r_s} + \left| \phi_{vsk} - \phi_{v'sk'} \right| / r_s + \left| \theta_{vsk} - \theta_{v'sk'} \right| / r_s \leq t_{vsk} \end{aligned} \tag{7}$$

$$t_{dsl}^b + T_g^{r_g} \leq t_{d's'l'}^a \quad \text{if} \quad t_{dsl}^b \leq t_{d's'l'}^a \tag{8}$$

$$t_{dsl}^b + T_s^{r_s} \leq t_{d's'l'}^a \quad \text{if} \quad t_{dsl}^b \leq t_{d's'l'}^a \tag{9}$$

$$I_s + \sum_{v \in \mathbf{V}} (\zeta_{vs} \cdot T_{vs}^p \cdot x_{vs}) - \sum_{d \in \mathbf{D}} \sum_{l \in \mathbf{L}_{ds}^{\mathbf{D}}} \gamma_s \cdot (t_{dsl}^b - t_{dsl}^a) \leq u_s \quad \text{for } t_{vsk}, t_{dsl}^a, t_{dsl}^b \leq t \quad^\forall t \in \{0, T_H\} \tag{10}$$

$$\left| \theta_{vsk} - \theta_{v_s sk_s} \right| \geq \beta_v \quad \text{if} \quad x_v = x_{v_s} = 1 \tag{11}$$

where,

$$\mathbf{x} = [x_v \ x_{v_s} \ x_{vs} \ x_{vsk}], \mathbf{t} = [t_{vsk} \ t_{dsl}^a \ t_{dsl}^b], \mathbf{\theta} = [\theta_{vsk}]$$

$$x_v, x_{v_s}, x_{vs}, x_{vsk} \in \{0,1\}, \quad \phi_{vsk} \in \{-\alpha_v, \alpha_v\} , \ \theta_{vsk} \in \{-\alpha_v, \alpha_v\} , \ t_{vsk}, t_{dsl}^a, t_{dsl}^b \in \{0, T_H\} ,$$

$$v, v_s, v' \in \mathbf{V}, d, d' \in \mathbf{D}, s, s' \in \mathbf{S}, k, k_s, k' \in \mathbf{K}_{vs}^{\mathbf{V}}, l, l' \in \mathbf{L}_{ds}^{\mathbf{D}}$$

$$v \neq v', d \neq d', s \neq s', k \neq k', l \neq l'$$

Eq. (1) is the objective function of the proposed problem, which is maximizing the sum of profits collected by conducting the selected tasks. Eq. (2) ~ Eq. (4) express the constraints on task allocation that specify that an observation task should be conducted only once (monoscopic) or only twice (stereoscopic). Indices for two observations related to the stereoscopic imaging task are $v$ are $v_s$. Note that only one of either Eq. (3) or Eq. (4) is active, depending on the type of task (monoscopic or stereoscopic). Eq. (5) expresses the constraint on the start time of observation tasks. It assigns a starting time value ($t_{vsk}$) within observation time window $k$ associated with an allocated task ($v$) and satellite ($s$) pair. Similarly, Eq. (6) ensures that all download tasks should occur within a download time window. Eq. (7) imposes a constraint that prevents unacceptable observation task overlaps in an observation time window. We can select two observation tasks simultaneously only if the first observation ends before the start of the second observation. For an acceptable task overlap, it is important to consider the transition time between two tasks, which is proportional to the sum of satellite stabilization time, difference in roll angles and difference in pitch angles. Note that, although the roll angle can be approximated as a constant for a given time window, the pitch angle varies within a range depending on the observation start time. The formulation adopts binary support variables to check the overlap between observation tasks. Eq. (8) and Eq. (9) describe the constraint to avoid overlaps in download tasks. Eq. (8) states that the next download of a ground station can start only after it finishes servicing a satellite and reorient the radar to serve another satellite. Similarly, Eq. (9) ensures that a satellite starts the next download (image data downlink) only after it finishes downloading for a ground station and contacts the next station. We assume that the preparation time for a ground station and a satellite is constant. Eq. (10) defines the constraint on the amount of data handled by a satellite. The amount of present data (= [accumulated observation data] - [accumulated download data]) cannot exceed the maximum satellite data storage capacity at all times during the scheduling time horizon. Eq. (11) imposes a constraint on stereoscopic imaging only. The pitch angle difference between the two observations for a stereoscopic imaging task should be larger than a pre-specified value.

## B.  Near-Optimal Heuristic for Satellite Task Scheduling

Direct implementation of the proposed MILP formulation could provide a true optimal solution. However, the number of decision variables for the MILP increases rapidly with the growth in key parameters determining the complexity of the problem (e.g., number of tasks, number of satellites, and number of ground stations). In addition, obtaining the exact solution of the problem becomes difficult when there are many time windows overlap constraints (regional and temporal concentration of imaging requests). Note that many time windows constraints associated with low-profit tasks are consequently inactive since a satellite will choose other higher-profit tasks, and are practically negligible. We propose a MILP-based heuristic that prunes out these constraints from the formulation and can yield near optimal solution within significantly reduced computation time.

The proposed MILP-based heuristic algorithm is composed of three steps. *Step* 1 identifies time window clusters that are potential bottlenecks and may consume large computation time. Time windows $k$ and $k'$ are grouped together in one time windows cluster if the gap between two time windows for a satellite is smaller than the maximum slew time ( $(2\phi_s^M + 2\theta_s^M)/r_s$ ) as follows

$$\left| T_{vsk}^b - T_{vsk'}^a \right| \leq (2\phi_s^M + 2\theta_s^M)/r_s, \ (k \neq k'; k, k' \in \mathbf{K}_{vs}^{\mathbf{V}}) \tag{12}$$

*Step 2* is an inner loop that sorts the time windows based on the priority values of their associated tasks. We leave $\lambda$ time windows with the 1) highest priorities and 2) lowest observation opportunities and remove the others from the cluster. The lowest observation opportunity (OP) is equal to the number of time windows related to task $v \in \mathbf{V}$, which we denote by $OP_v$. We suggest the lower bound of the parameter ($\lambda_{LB}$) obtained by dividing the *average length of observation time windows* by the sum of *average observation time* and *satellite stabilization time* as follows:

$$\lambda_{LB} = \left\lceil \frac{\left( \sum_{v \in \mathbf{V}} \sum_{s \in \mathbf{S}} \sum_{k \in \mathbf{K}_s^{\mathbf{V}}} (T_{vsk}^b - T_{vsk}^a) \right)/N_{tw}}{\left( \sum_{v \in \mathbf{V}} \sum_{s \in \mathbf{S}} T_{vs}^p \right)/N_{task} + \left( \sum_{s \in \mathbf{S}} T_s^{r_s} \right)/N_{sat}} \right\rceil \tag{13}$$

where $N_{tw}$ is the number of time windows, $N_{task}$ the number of tasks, and $N_{sat}$ is the number of satellites. When the priorities and observation opportunities of multiple time windows are same, the heuristic chooses the group of windows with the smallest deviation from the average roll angle of already selected high-priority windows. Windows with similar roll angles result in short cross-range distance, which leads to the reduction of transition time between tasks and potential increase in number of observation tasks. If $n$ higher priority windows are already selected $(n < \lambda)$, the heuristic selects the next $(\lambda - n)$ windows in descending priority order whose roll angles are closest to the average roll angles of the already selected $n$ higher priority windows as

$$v = \arg\max \delta_{vsk} = \left| \phi_{vsk} - (\sum_n \phi_{v'sk'})/n \right| \tag{14}$$

where $v \in V, v' \in V', s \in S, k \in \mathbf{K}_{vs}^{\mathbf{V}}, k' \in \mathbf{K}_{v's}^{\mathbf{V'}}$, and $V'$ is the set of already selected high-priority windows. *Steps* 1-2 run iteratively until all existing clusters have only $\lambda$ high-priority time windows.

*Step* 3 rearranges the time windows that are not pruned out (after *Step* 2), creates the modified MILP, and solves the problem. Solving the modified MILP yields near-optimal schedule while significantly reducing the computation time, which is validated through the case study in the next section. Table 4 presents the pseudocode of the proposed MILP-based heuristic, which is visualized in Figure 6 as well. The numbers inside the boxes of the figure represent the task priority values for observation tasks.

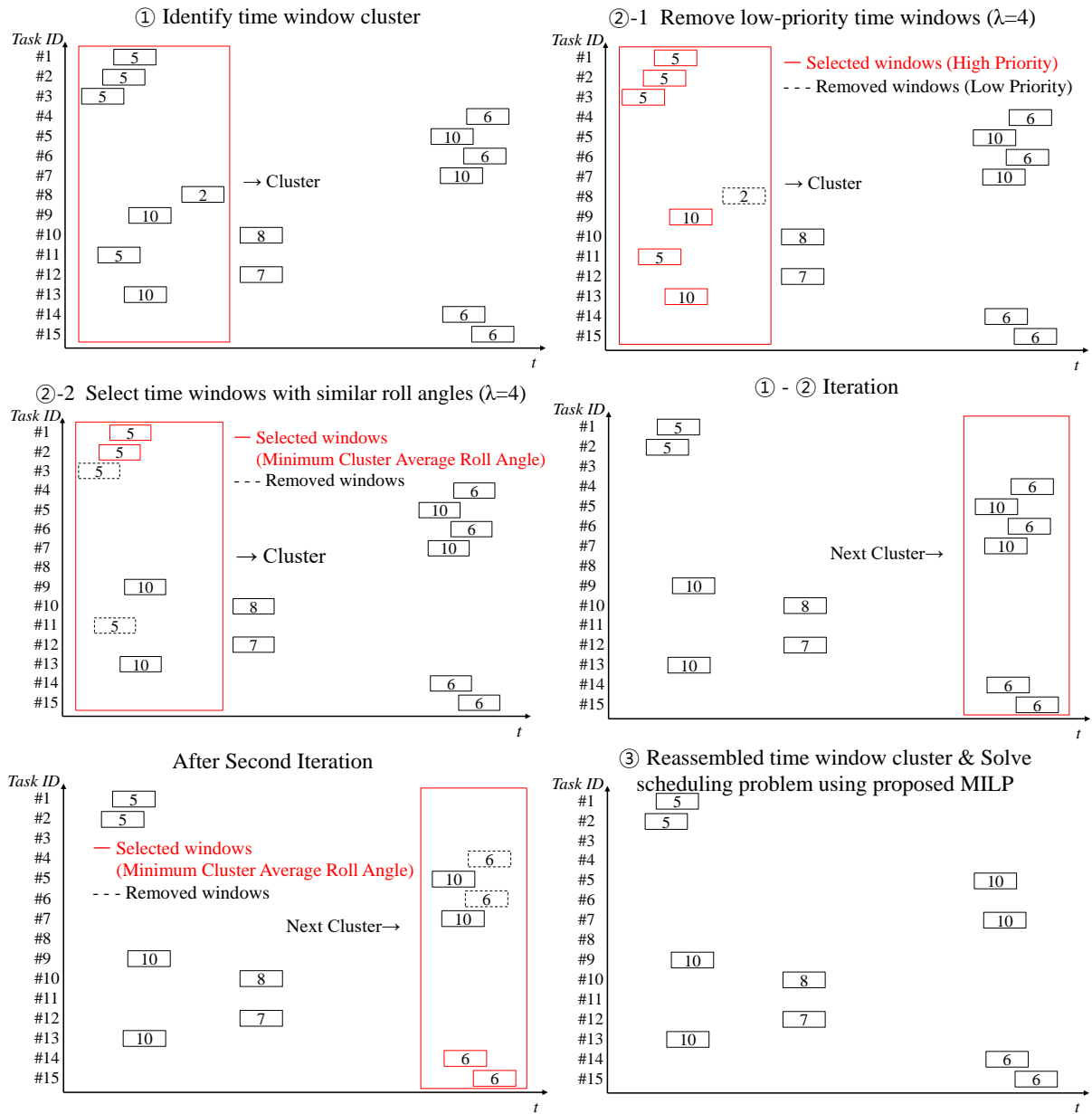

**Figure 6. Steps of MILP-based Heuristic in a Sample Cluster ($\lambda$=4)**

**Table 1. MILP-based Heuristic Pseudocode for satellite task scheduling**

|  |  |  |
|---|---|---|
|  | 1 | **Receive** $v \in \mathbf{V}, s \in \mathbf{S}, k \in \mathbf{K}_{vs}^{\mathbf{V}}$, Breaker = 0 |
| *Step* 1 | 2 | **While** Breaker<1 |
|  | 3 | Count = 0 |
|  | 4 | **Sort** $k$ by $t_{vsk}$ |
|  | 5 | For $k \neq k'$, add $k$ to a **Cluster** |
|  | 6 | **If** $T_{vsk}^b + (2\phi_s^M + 2\theta_s^M)/r_s \leq T_{vsk}^a$. and $T_{vsk}^a \leq T_{vsk}^a$. |
| *Step* 2 | 7 | **While** $N > \lambda$ where $N = \#k \in \mathbf{Cluster}$, $n \in N$ |
|  | 8 | **Sort** $k_n$ by max $w_v$ then by min $OP_v$ |
|  | 9 | For equal $w_v$, Sort by min $\delta_{vsk}$ |
|  | 10 | **For** $k_n > \lambda$ in a cluster |
|  | 11 | **If** $\sum_s k \in \mathbf{K}_{vs}^{\mathbf{V}} \geq 1$ |
|  | 12 | **Remove** $k_n$ |
|  | 13 | **End** |
|  | 14 | **Sort** $k_n$ by $t_{vsk}$ |
|  | 15 | **End** |
|  | 16 | Count += 1 |
|  | 17 | **If** Count>Total #k |
|  | 18 | Breaker = 1 |
|  | 19 | **End** |
|  | 20 | **Return** $v \in \mathbf{V}, s \in \mathbf{S}, k_n \in \mathbf{K}_{vs}^V$ |
| *Step* 3 | 21 | **Solve** schedule using proposed MILP formulation |

## IV. Case Study

### A. Introduction

A case study has been conducted to demonstrate the effectiveness of the MILP formulation and the heuristic algorithm proposed in this paper. Eight problem instances (6 spot-imaging and 2 strip-imaging missions) with 50 – 100 observation tasks in East Asia were created and solved using the proposed approach. The observation tasks and ground stations were randomly distributed spatially and the time windows (start/end times) and pitch/roll angles required to conduct the tasks were computed as a pre-processing. The values representing the priority were randomly (uniform distribution) generated and assigned to the tasks. Figure 7 shows the task locations plotted in the global map for one of problem instances used for the case study.

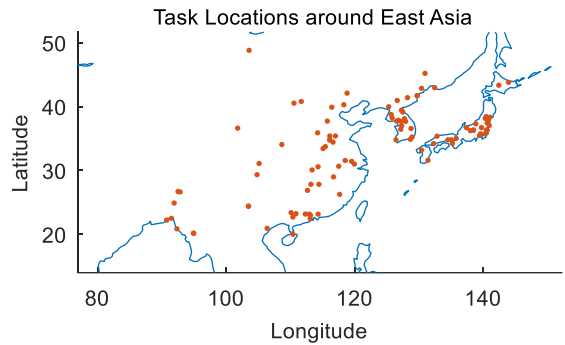

**Figure 7. Instance 100-Cp-1, 100 tasks around East Asia**

Table 2 shows the orbital parameters of the satellites used for the mission. Some parameters such as task observation time, satellite stabilization time, and required roll angle are determined depending on the task and the satellite conducting the task, which are generated based on the orbital simulation. Note that as $\lambda$ approaches infinity, the solution obtained by the MILP-based heuristic gets closer to the true optimal. Figures 8 - 10 show the task locations plotted in the global map for some of the generated instances and Table 3 presents the scheduling parameters.

**Table 2. Orbital parameters of satellites**

| Satellite ID | $a$, km | $e$, - | $i$, deg | $\omega$, deg | RAAN, $deg$ |
|---|---|---|---|---|---|
| Satellite #1 | 6871 | 0 | 97.3 | 0 | 0 |
| Satellite #2 | 6871 | 0 | 97.3 | 90 | 90 |
| Satellite #3 | 6871 | 0 | 97.3 | 180 | 180 |
| Satellite #4 | 6871 | 0 | 97.3 | 270 | 270 |

**Table 3. Scheduling Parameters**

| Parameter | Value | Parameter | Value |
|---|---|---|---|
| $T_H$ , hr | 24 | $T_{vsk}^p$ , s | 3 (Spot) / 15 (Strip) |
| $w_v$ , - | 1-5 (uniform distribution) | $T_s^{r_s}$ , s | 5 |
| $\phi_{vsk}$ , - | Calculated | $T_g^{r_g}$ , s | 60 |
| $\alpha_v$ , | Not designated | $T_s^{r_g}$ , s | 20 |
| $\phi_s^M ; \theta_s^M$ , deg ; deg | 30 | $\beta_v$ , deg | 15 |
| $r_s$ , deg/s | 1 | $\gamma_s$ , - | 5 |
| $T_{vsk}^a$ , s; $T_{vsk}^b$ , s | Calculated | $\lambda_{low}$ , - | 12 |
| $T_{dsl}^a$ , s; $T_{dsl}^b$ , s | Calculated | | |

The implementation and solution of the proposed formulation (direct MILP and MILP-based heuristic) were conducted using the Python language and the GUROBI solver with Intel Core i7 processor with 16 GB of memory. Performances of the solution methodologies introduced in this study along with the results obtained using the first-in-first-out (FIFO) heuristic are presented in Table 4. In the table, *relative performance* represents the value of objective function compared to that of the direct MILP implementation. The maximum computation time of the direct MILP solution was set up as 3 hr. If the solution does not converge until this time, the optimizer provides the best solution found so far along with the worst-case optimality gap between the true optimal and obtained solutions.

We can observe that, while the performance of MILP based heuristic is quite close to that of the exact MILP approach, its computation time is significantly shorter. For instances with 100 tasks, when the $\lambda$ value is small (e.g., $\lambda$ = 12), the MILP heuristic obtained 90 ~ 100 % of the performance (= objective function) of the direct MILP while using only 0.09 ~ 4.76 % of computation time (average: 1.28 %). With a large $\lambda$ value (e.g., $\lambda$ = 15), the MILP heuristic performs closer to the direct MILP (97 ~ 100%) while consuming 0.23 ~ 9.22 % of computation time (average: 3.39 %). The aforementioned results imply the trade-off between the performance and the computation time: the larger $\lambda$ value corresponds to the better solution quality and the longer computation time. The performance of the FIFO heuristic is relatively low (62 ~ 83 % of the direct MILP results) while its computation time is very short compared to other two methods. We can conclude that the attractiveness of the FIFO heuristic is relatively low unless the situation requires very urgent scheduling (order of seconds), which usually not the case for scheduling of satellite constellation.

**Table 4. Performances: exact MILP solution, MILP-based heuristic (various λ values), and FIFO**

| Scenario | Algorithm | Obj. Function (J), -* | # Assigned Tasks, - | Relative Performance, % | Computation Time, s | Rel. Time Consumption, % |
|---|---|---|---|---|---|---|
| 50-Cp-1 | Direct MILP | 214 | 33 | 100 | 12 | 100 |
| | MILP-heuristic (12) | 214 | 33 | 100 | 12 | 100 |
| | MILP-heuristic (13) | 214 | 33 | 100 | 12 | 100 |
| | MILP-heuristic (14) | 214 | 33 | 100 | 12 | 100 |
| | MILP-heuristic (15) | 214 | 33 | 100 | 12 | 100 |
| | FIFO | 178 | 28 | 83 | 0.3 | 2.5 |
| 100-Cp-1 | Direct MILP | 394 | 61 | 100 | 13566 | 100 |
| | MILP-heuristic (12) | 362 | 53 | 92 | 12 | 0.09 |
| | MILP-heuristic (13) | 374 | 55 | 95 | 18 | 0.13 |
| | MILP-heuristic (14) | 378 | 56 | 96 | 26 | 0.19 |
| | MILP-heuristic (15) | 386 | 58 | 98 | 31 | 0.23 |
| | FIFO | 256 | 44 | 65 | 2 | 0.01 |
| 100-Cp-2 | Direct MILP | 332 (1%) | 57 | 100 | 86540 | 100 |
| | MILP-heuristic (12) | 326 | 53 | 98 | 739 | 0.85 |
| | MILP-heuristic (13) | 328 | 54 | 99 | 997 | 1.15 |
| | MILP-heuristic (14) | 330 | 55 | 99 | 1183 | 1.37 |
| | MILP-heuristic (15) | 332 | 56 | 100 | 2933 | 3.39 |
| | FIFO | 218 | 41 | 66 | 1 | 0.00 |
| 100-Cp-3 | Direct MILP | 378 (1.6%) | 60 | 100 | 86481 | 100 |
| | MILP-heuristic (12) | 340 | 50 | 90 | 313 | 0.36 |
| | MILP-heuristic (13) | 352 | 52 | 93 | 1065 | 1.23 |
| | MILP-heuristic (14) | 362 | 56 | 96 | 1617 | 1.87 |
| | MILP-heuristic (15) | 366 | 57 | 97 | 862 | 1.00 |
| | FIFO | 234 | 40 | 62 | 2 | 0.00 |
| 100-Cp-4 | Direct MILP | 458 (9.6%) | 78 | 100 | 86444 | 100 |
| | MILP-heuristic (12) | 406 | 64 | 90 | 715 | 0.83 |
| | MILP-heuristic (13) | 420 | 68 | 92 | 1281 | 1.48 |
| | MILP-heuristic (14) | 434 | 70 | 95 | 2739 | 3.17 |
| | MILP-heuristic (15) | 442 | 71 | 97 | 1696 | 1.96 |
| | FIFO | 284 | 52 | 62 | 3 | 0.00 |
| 100-Cp-5 | MILP | 344 | 75 | 100 | 68140 | 100 |
| | MILP-heuristic (12) | 328 | 68 | 95 | 3239 | 4.75 |
| | MILP-heuristic (13) | 328 | 68 | 95 | 3794 | 5.57 |
| | MILP-heuristic (14) | 332 | 69 | 97 | 5319 | 7.81 |
| | MILP-heuristic (15) | 336 | 71 | 98 | 4491 | 6.59 |
| | FIFO | 224 | 56 | 65 | 3 | 0.00 |
| 100-Ct-1 | MILP | 310 (12%) | 45 | 100 | 86501 | 100 |
| | MILP-heuristic (12) | 284 | 40 | 92 | 354 | 0.41 |
| | MILP-heuristic (13) | 292 | 41 | 94 | 300 | 0.35 |
| | MILP-heuristic (14) | 304 | 43 | 98 | 711 | 0.82 |
| | MILP-heuristic (15) | 304 | 43 | 98 | 1180 | 1.36 |
| | FIFO | 234 | 41 | 83 | 1 | 0.00 |
| 100-Ct-2 | MILP | 270 | 42 | 100 | 38492 | 100 |
| | MILP-heuristic (12) | 266 | 41 | 99 | 640 | 1.66 |
| | MILP-heuristic (13) | 268 | 42 | 99 | 1438 | 3.74 |
| | MILP-heuristic (14) | 270 | 42 | 100 | 1941 | 5.04 |
| | MILP-heuristic (15) | 270 | 43 | 100 | 3550 | 9.22 |
| | FIFO | 218 | 41 | 81 | 1 | 0.00 |

* The number in the parenthesis in this column represents the optimality gap for an instance whose direct MILP solution does not converge.

## V.     Conclusions

The proposed MILP formulation for optimal scheduling of Earth imaging by multiple agile satellites effectively reflects various practical considerations. Because task requests frequently exceed the process capability of the satellites, the formulation maximizes the sum of profit values (representing the priority) allocated to the observation tasks while considering attitude transitions, downloading to ground stations, satellite data storage capacity, task overlaps, and time windows.

Due to the inherent complexity of the formulated problem, obtaining its exact solution by directly solving the MILP often requires computational time that is too long to be considered as a practical runtime algorithm. A heuristic algorithm to find a near-optimal solution of the original MILP is proposed to address its issue on the long computation time. The heuristic prunes out irrelevant design variables and associated time windows constraints through a priority-based time window sorting technique. A case study on realistic satellite image scheduling demonstrates that the developed MILP formulation provides solutions applicable for practical satellite operations and the proposed heuristic obtains near-optimal solutions (2 ~ 8 % worst-case error compared with true optimal depending on the value of parameter $\lambda$ ) while accelerating the computation significantly.

Potential subjects for future study include the application of the proposed approach on problems with larger task numbers and/or various task types (e.g., polygon that requires many strips for full area coverage). Development of scheduling adjustment methodology that can address the emergent situations within the scheduling horizon can be also an interesting subject as well.